\documentclass[a4paper]{jpconf}
\usepackage{graphicx}
\usepackage{hyperref}
\hypersetup{
    colorlinks=true,       
    linkcolor=black,          
    citecolor=black,        
    filecolor=black,      
    urlcolor=black           
}
\begin{document}

\title{GPU-accelerated large-scale quantum molecular dynamics simulation of 3-dimensional $\mathrm{C}_{60}$ polymers}

\author{Toshiaki Iitaka}

\address{Computational Astrophysics Laboratory, RIKEN \\ 
2-1 Hirosawa, Wako, Saitama 351-0198, Japan \\
\url{http://www.iitaka.org/}
}

\ead{tiitaka@riken.jp}

\begin{abstract}
Polymerization of $\mathrm{C}_{60}$ molecular crystal under high pressure and high temperature is simulated by using linear scaling tight binding molecular dynamics (TBMD) with Graphic Processing Unit (GPU) as a computational accelerator for matrix-matrix multiplication. Two sets of tight binding parameters were tested.
\end{abstract}

\section{Introduction}
High pressure transformation of $\mathrm{C}_{60}$ molecular crystal has been attracting a great deal of attention in the past years because of its potential superconductivity and super-hardness. At ambient conditions, $\mathrm{C}_{60}$ molecules crystallize into a face-centered-cubic (fcc) structure with weak van der Waals interactions. Under extreme compression, the fcc $\mathrm{C}_{60}$ crystal may transform into polycrystalline diamond, diamond-like materials, or graphitic forms. Yamanaka et al. \cite{Chen2002,Yamanaka2006,Yamanaka2008} have succeeded in synthesizing two dimensional and three dimensional $\mathrm{C}_{60}$ polymers at moderate pressure and temperature by forming covalent bonds between $\mathrm{C}_{60}$ molecules via hybridization change of $sp^2$ bond to $sp^3$ bond. Several polymorphs of 3-d $\mathrm{C}_{60}$ have been theoretically proposed \cite{Yamanaka2008,Yang2007a,Yang2007b,Zipoli2008,Yamagami2009} and studied by using electronic structure calculation, geometry optimization and phonon calculation. However, the experimentally observed structure is not fully consistent with any of theoretically predicted models. Therefore molecular dynamics simulation of polymerization of $\mathrm{C}_{60}$ is important to understand the crystal structure. In this paper we introduce a large scale tight binding molecular dynamics simulation of 3-d $\mathrm{C}_{60}$ polymer formation with thousands of carbon atoms in the unit cell. Such large scale quantum simulation was achieved by the combination of linear scaling tight-binding molecular dynamics (TBMD) \cite{Takayama2004,Takayama2006,Hoshi2009,Ordejon1993,Ozaki2006} and computational acceleration with graphic processing unit (GPU) \cite{CUDA,Iitaka2007,Hamada2007,HaRiken09}.

\section{Linear scaling method}
The Hamiltonian of non-selfconsistent field tight binding method is written as
\begin{equation}
\label{eq:Hamiltonian}
 H = \sum_{i,\alpha} \epsilon_{i,\alpha} c^{\dagger}_{i,\alpha} c_{i,\alpha}
+\sum_{i,\alpha,j,\beta} t_{i,\alpha; j,\beta} c^{\dagger}_{i,\alpha} c_{j,\beta}
\end{equation}
where $i$ and $j$ are the indices of atoms and $\alpha$ and $\beta$ are the indices of orbitals of the atom while $c^{\dagger}$ and $c$ are the creation and annihilation operators of electron.

The one-body density matrix of the Hamiltonian (\ref{eq:Hamiltonian}) is defined as \begin{equation}
\rho = \sum_{m} \left| m \rangle f(\epsilon_m) \langle m \right|
\end{equation}
where $| m \rangle$ are the eigenstates of (\ref{eq:Hamiltonian}) and $f(\epsilon_m)$ is the occupation number of the eigenstate $| m \rangle$.  The density matrix contains all information of the system state. For example, the force on the $i$-th atom can be calculated as $\langle F_{i} \rangle = \Tr[\rho F_{i}]$ by using the density matrix and the force matrix. The computational cost is $O(N^3)$ due to the diagonalization of Hamiltonian where N is the total number of atoms.

Linear scaling of the computational cost with respect to $N$ is achieved by exploiting the locality of the density matrix, or solving the local electronic properties of the target atom by taking into account the atoms within the cutoff radius from the target atom only \cite{Takayama2004, Takayama2006, Hoshi2009, Ordejon1993, Ozaki2006}. Then the computational cost becomes $O(n^3N)$ where $n$ is the number of atoms within the cutoff radius from the target atom.


\begin{figure}
\begin{center}
\includegraphics[width=28pc]{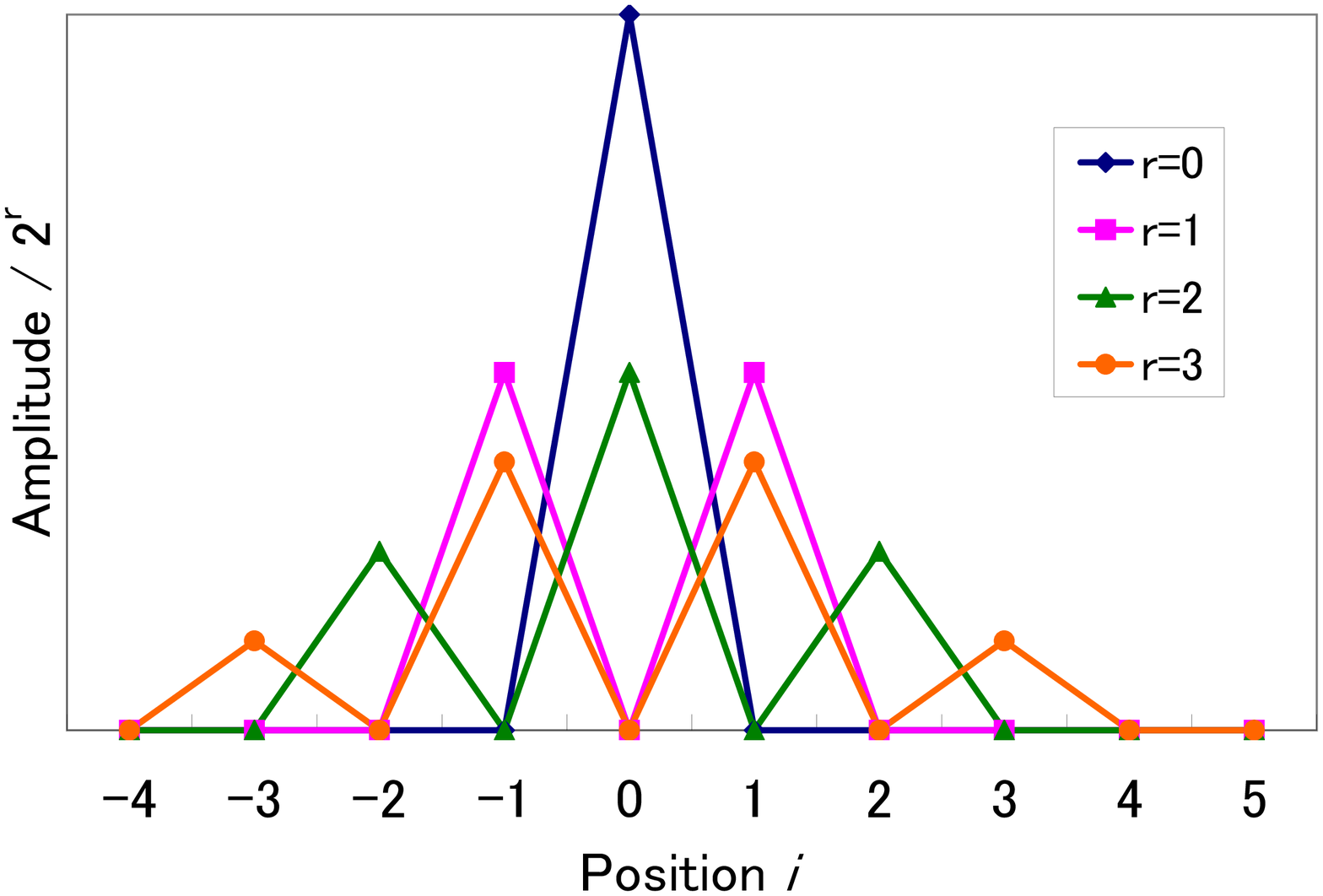}
\end{center}
\caption{\label{fig:1} Krylov basis of order $r$  with the seed vector $|i=0\rangle$ for one dimensional uniform chain.}
\end{figure}

\section{Krylov subspace}
The Hilbert space of the Hamiltonian of the atoms within the cutoff radius from the target atom is expressed as
\begin{equation}
\mathrm{span} \left\{ \left|1,\alpha\right>, \left|2,\alpha\right>, \left|3,\alpha\right>,\cdots \left|i,\alpha\right> \cdots, \left|n,\alpha\right> \right\}
\end{equation}
where $n$ is the number of the atoms in the cutoff sphere and $\alpha$ is the index of the atomic orbital, which is hereafter omitted for simplicity. The Krylov subspace of order $r$ with the {\it seed vector} $|\phi\rangle$, $K_r(H,|\phi\rangle)$, is generated by multiplying $H$ repeatedly on the seed vector $|\phi\rangle$. 
\begin{equation}
\label{eq:Krylov}
K_r(H,|\phi\rangle)=\mathrm{span} \left\{ \left|\phi\right>, H\left|\phi\right>, H^2\left|\phi\right>,\cdots ,H^{r-1}\left|\phi\right>\right\}.
\end{equation}
Here we use the atomic orbital of the $i$-th atom as the seed vector, $|\phi\rangle=|i\rangle$, to calculate the force on the $i$-th atom.  This is quite different from the usual Lanczos diagonalization methods where the random vector is used for the seed vector.
The accuracy of the Krylov subspace as an approximant of the original Hilbert space increases rapidly as $r$ increases up to $m$, where $m < n$ is the degree of the minimal polynomial of $H$ with respect to $|i\rangle$. It was shown \cite{Takayama2004} that Krylov subspace of order $r \ll n$ can describe the system with sufficient accuracy. The efficiency of the Krylov subspace can be understood from the physical point of view.  Since the tight binding Hamiltonian (\ref{eq:Hamiltonian}) contains the hopping terms $V_{ij}$, multiplication of $H$ on $|i\rangle$ makes the electron hop from the $i$-th atom to the neighbor atoms linked by the hopping terms. Multiplying $H$ twice on $|i\rangle$ makes the electron spread to the next nearest neighbors. Therefore the Krylov basis generated from the seed vector $|i\rangle$ is localized at the target atom and then gradually spreads out to nearby atoms. Let us examine this with the simplest example of one dimensional uniform chain whose Hamiltonian matrix elements (\ref{eq:Hamiltonian}) are all zero except $t_{i,i\pm1}=1$ for all ${i}$.  Then the Krylov basis of order $r$ can be expressed in terms of binomial coefficients,
\begin{equation}
H^{r}|i=0\rangle=\sum_{k=0}^{r} C_{r,k}|2k-r\rangle.
\end{equation}
In Fig.~\ref{fig:1} the Krylov basis in real space representation is shown.
Krylov subspace of small order ($r \approx 30$) can represent quantum states localized around the target atom accurately because it samples the wave function around the atom with larger weight.

\begin{figure}
\begin{center}
\includegraphics[width=28pc]{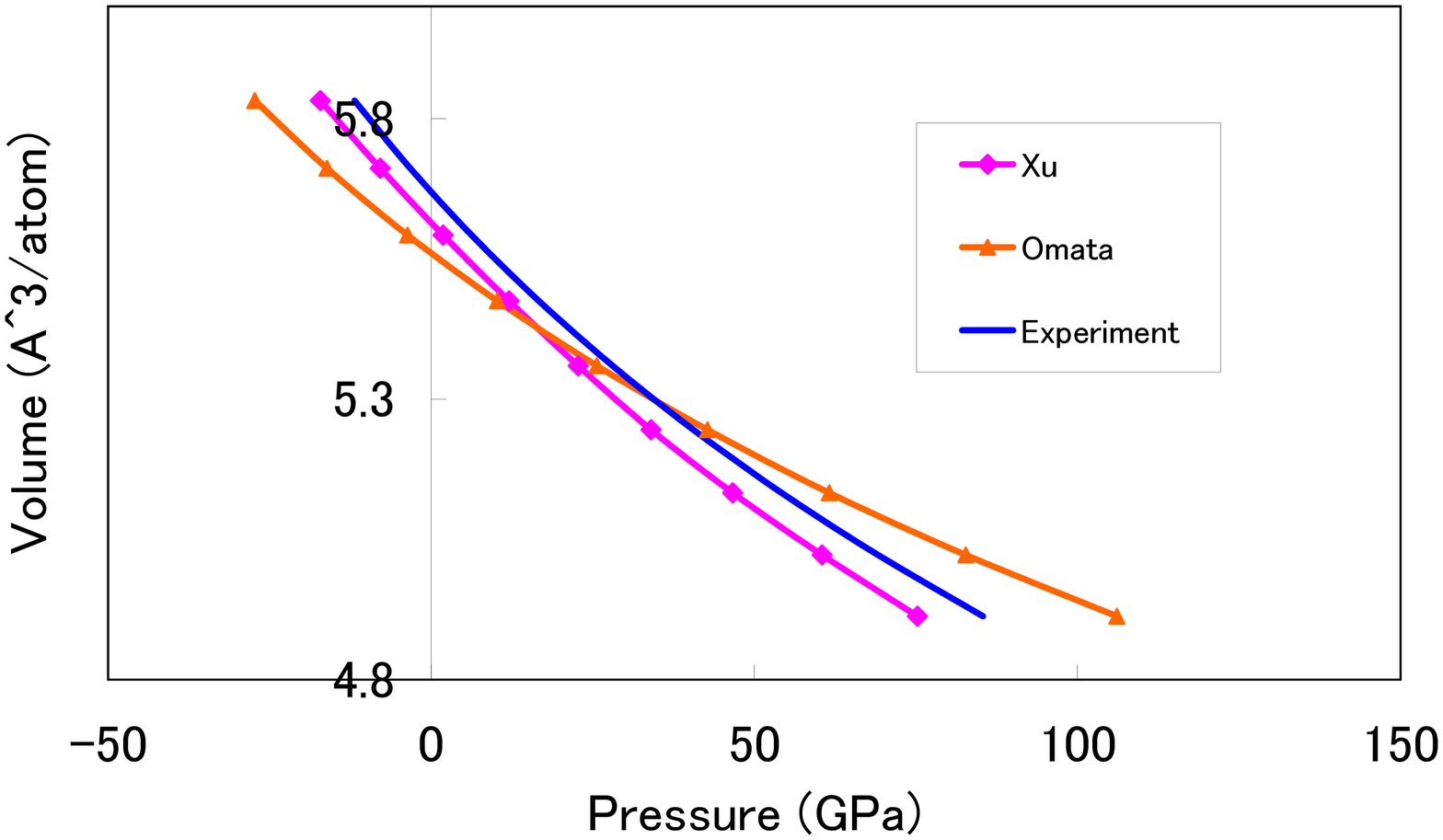}
\end{center}
\caption{\label{fig:2} Equations of state of the cubic diamond: (1) calculated data with Xu's parameter \cite{Xu1992}, (2) calculated data with Omata's parameter \cite{Omata2005}, (3) experimental data by Dewaele et al. \cite{Dewaele2008}.}
\end{figure}

Since Krylov basis (\ref{eq:Krylov}) is not orthogonal, Lanczos algorithm is often used to construct orthogonal basis set from it. 
Lanczos algorithm transforms an Hermitian matrix $H$ to a tridiagonal real symmetric matrix $T=V^\dagger H V$ where the transform matrix $V=\left( v_1, v_2,\cdots, v_r \right)$ consists of orthogonal basis $v_{m}$ called Lanczos vector.
The Lanczos vector $v_{m}$ and the diagonal and off-diagonal matrix elements of $T$ ($\alpha_i$ and $\beta_i$) are calculated by the three term recursion formula:
\begin{eqnarray}
\label{eq:lanczos}
\left| l_k \right\rangle &=& H \left|v_k\right\rangle -\alpha_k \left|v_k\right\rangle -\beta_{k-1} \left|v_{k-1} \right\rangle \\
\left|v_{k+1}\right\rangle &=& \frac{1}{\beta_k} \left| l_k \right\rangle 
\end{eqnarray}
where $v_{0}=0$,$v_{1}=|i\rangle$,$\alpha_k=\langle v_k| H| v_k \rangle$, $\beta_k=\sqrt{\langle l_k| l_k \rangle}$ and $\beta_0=0$.
Once $T$ is obtained, the eigenvalues and eigenvectors of this small sized ($r\times r$) symmetric tridiagonal matrix are easily calculated by using common linear algebra libraries such as LAPACK. 

In order to calculate all forces acting on the N atoms we have to repeat the above procedure for each atom. By rearranging the atoms into small groups consisting of $m$ atoms ($m \ll n \ll N$) and calculating simultaneously the forces acting on the $m$ atoms, we can rewrite the three term recursion formula (\ref{eq:lanczos}) in the matrix-matrix multiplication form (Level 3 BLAS form) \cite{BLAS}, which is efficiently calculated with GPU.
\begin{eqnarray}
\label{eq:lanczos_blas3}
L_k     &=&  H V_k - A_k V_k -B_{k-1} V_{k-1} \\
V_{k+1} &=&  B_k^{-1} L_k 
\end{eqnarray}
where $L_k=(| l_k^{(1)}\rangle, | l_k^{(2)} \rangle,\cdots,| l_k^{(m)} \rangle)$, $V_k=(v^{(1)}_k,v^{(2)}_k,\cdots,v^{(m)}_k)$, $A_k$ is diagonal matrix $[\alpha_k^{(1)},\alpha_k^{(2)},\cdots,\alpha_k^{(m)}]$, and $B_k$ is diagonal matrix $[\beta_k^{(1)},\beta_k^{(2)},\cdots,\beta_k^{(m)} ]$. 

\begin{figure}
\begin{minipage}{36pc}
\begin{center}
\includegraphics[width=18pc]{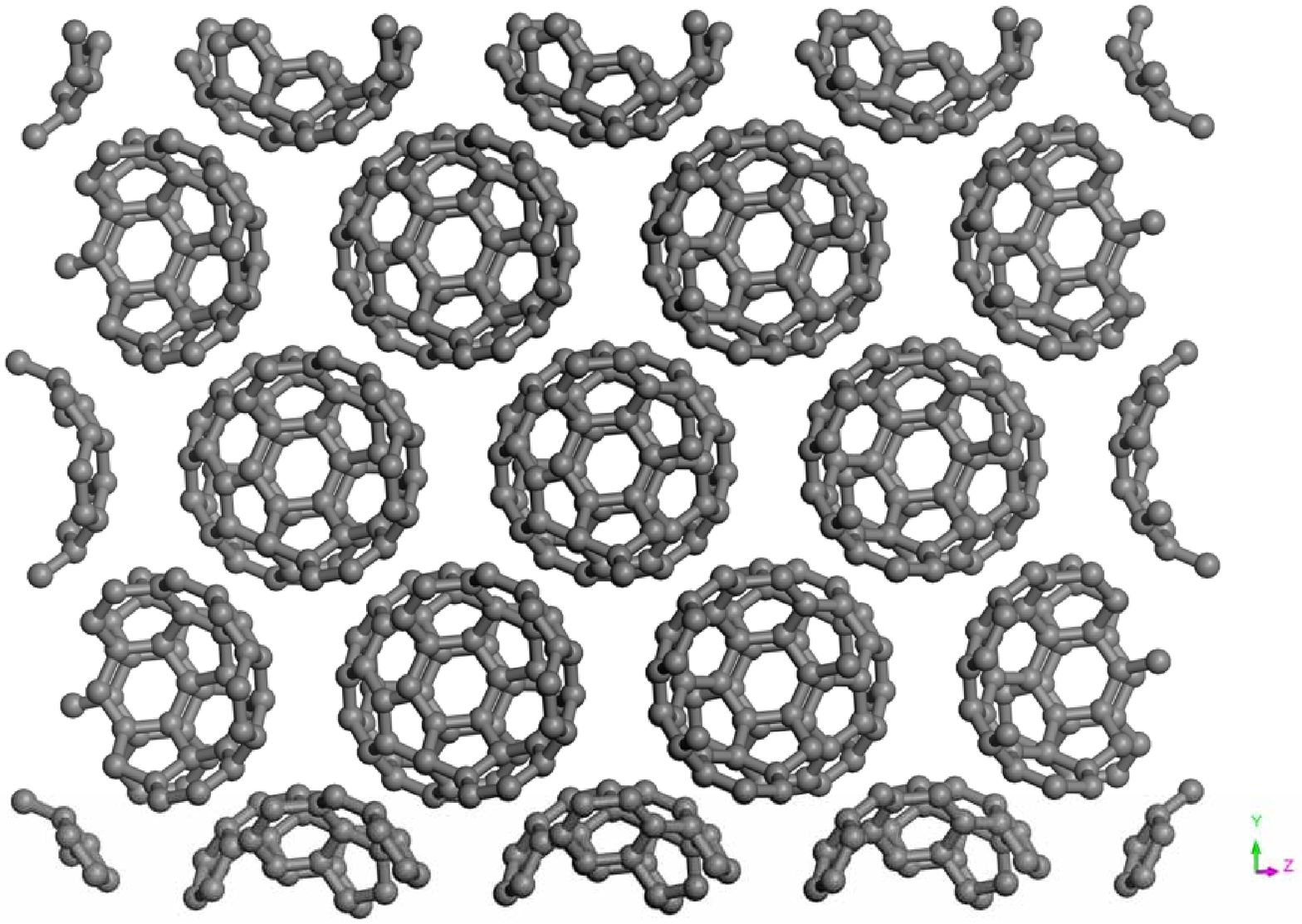}
\end{center}
\caption{\label{fig:3} the 2x2x2 fcc structure of the $\mathrm{C}_{60}$ at 0GPa and 0K.} 
\end{minipage}\hspace{2pc}%

\begin{minipage}{36pc}
\begin{center}
\includegraphics[width=18pc]{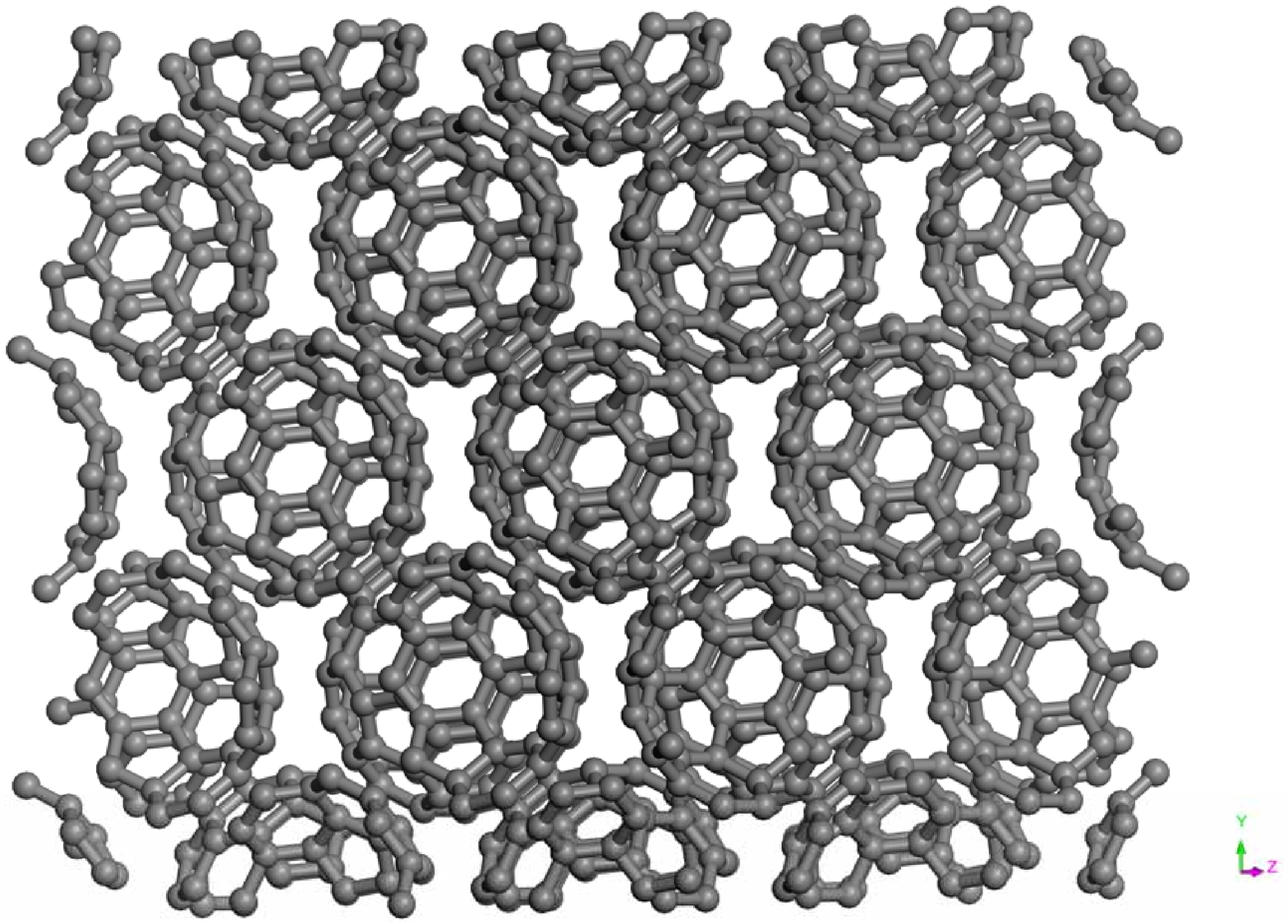}
\end{center}
\caption{\label{fig:4} 
Snapshot after 3 ps NPT simulation with Xu's parameter at 25 GPa and 500K.}
\end{minipage}\hspace{2pc}%

\begin{minipage}{36pc}
\begin{center}
\includegraphics[width=18pc]{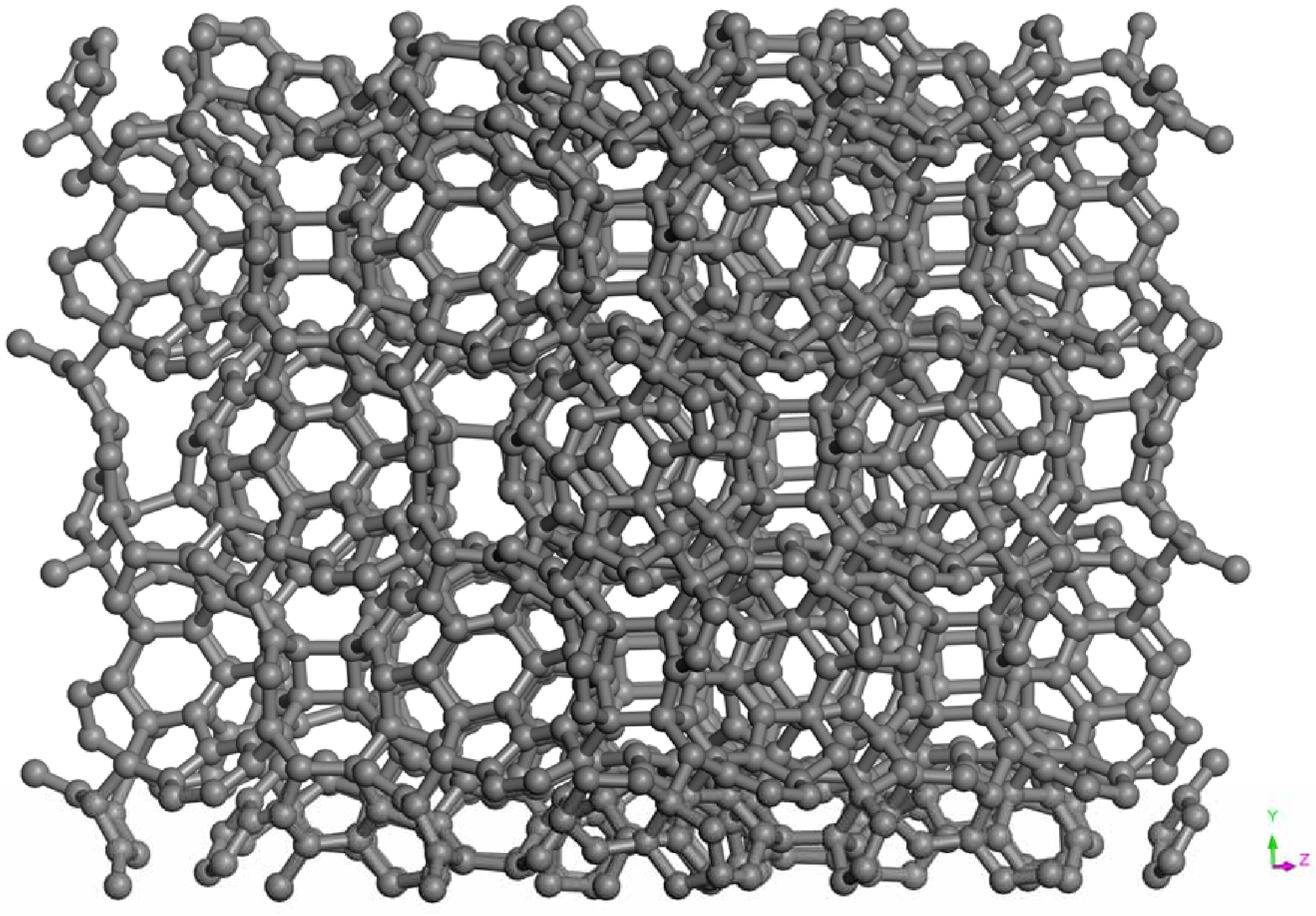}
\end{center}
\caption{\label{fig:5} 
Snapshot after 3 ps NPT simulation with Omata's parameter at 25 GPa and 500K.}
\end{minipage} 

\end{figure}

\section{Molecular dynamics of $\mathrm{C_{60}}$}
The tight binding method was applied to the equation of state (EOS) of diamond and molecular dynamics of $\mathrm{C}_{60}$ polymerization.  For this purpose, two types of tight binding model parameters, Xu's parameter\cite{Xu1992} and Omata's parameter\cite{Omata2005} were used. Both parameter sets have the same functional forms, but Omata's parameter was constructed to reproduce the LDA energetics of not only the $sp^2$ graphene and $sp^3$ diamond covalent bonds but also that of the $sp^2$ interlayer interaction. The cutoff radius of Xu's parameter is 2.6 A, while that of Omata's parameter is 7.0 A so that it can include interaction between $sp^2$ interlayers, which may be important for polymerization of $\mathrm{C}_{60}$.

The EOS of 4x4x4 supercell of cubic diamond was calculated for Xu's and Omata's TB parameters (Fig.~\ref{fig:2}). Convergence was tested with respected to the number of atoms in the cutoff radius $n$, the order of Krylov subspace $r$ and size of supercell (by using 8x8x8 supercell). The EOS of Xu's TB parameter reproduces the experimental result well while the Omata's parameter produces stiffer EOS.

Fig.~\ref{fig:3} shows the 2x2x2 fcc structure of the $\mathrm{C}_{60}$ at 0K and 0GPa as the initial structure, Fig.~\ref{fig:4} shows the snapshot after 3 ps NPT simulation with Xu's parameter at 25 GPa and 500K, and Fig.~\ref{fig:5} shows the snapshot after 3 ps NPT simulation with Omata's parameter at 25 GPa and 500K.  Simulation with Omata's parameter successfully reproduced the polymerization of $\mathrm{C}_{60}$ under high temperature and high pressure. However, the produced polymer was not perfect crystal. The finer control of temperature and pressure would be required to obtain better results.

\section{Acceleration by GPU}
Graphics Processing Unit (GPU) is an electronic device for accelerating graphic processing on personal computers and game machines. GPU can perform numerical calculation much faster than CPU in some types of computation such as matrix-matrix multiplication \cite{CUDA}, fast Fourier transformation (FFT) \cite{CUDA}, and the N-body problem \cite{Hamada2007}. Application of GPU computation is rapidly expanding to many fields of science including brain science, quantum science, astronomy and fluid dynamics \cite{HaRiken09}. In this paper, we apply GPU computation to tight binding molecular dynamics.
According to the Amdal's law \cite{Amdal} the speedup of a program using GPU is limited by the time needed for the CPU fraction of the program. The breakdown of computational time of the original CPU version of ELSES \cite{Hoshi2009} shows that the most time consuming part is the matrix-vector multiplication in the Lanczos algorithm which is 99.92 \% of the total computation time in our case. Therefore the calculation can be significantly accelerated by using fast matrix-vector operation.  For this purpose, we adopted CUBLAS, a linear algebra library on GPU \cite{CUDA}.
The elapse time for calculating the pressure of 4x4x4 supercell of diamond with Omata's parameter was 109.53 $s$ for GPU version with $m=4$ and 504.3 $s$ for the original CPU version of ELSES \cite{Hoshi2009}. Therefore the GPU version is five times faster than the CPU version. The elapse time for Xu's parameter was 55.65 $s$ for GPU version and 29.19 $s$ for CPU version. The CPU calculation is efficient when the interaction is short ranged and $H$ becomes sparse matrix, while GPU becomes very efficient when the interaction is long ranged and the Hamiltonian becomes dense matrix.
We used Sun Fire X2250 eight core workstation for CPU computation and NVIDIA GeForce GTX8800 for GPU computation.


In summary, formation process of $\mathrm{C}_{60}$ polymer under high pressure and high temperature was simulated by the linear scale tight binding molecular dynamics with hardware acceleration by GPU. Two sets of tight binding parameters were tested. Xu's parameter has short ranged interaction resulting in fast simulation on CPU but does not faithfully reproduce polymerization. Omata's parameter has long ranged interaction and enjoy the acceleration by GPU. It reproduces $sp^2$-$sp^3$ hybridization change successfully but does not reproduce experimental EOS accurately. Further development of accurate tight binding parameters for $\mathrm{C}_{60}$ polymer would be necessary.

This research was supported partially by Grant-in-Aid for Scientific Research on Innovative Areas 'Earth science based on the high pressure and temperature neutron experiments' (No. 20103001-20103005), from the Ministry of Education, Culture, Sports, Science and Technology (MEXT) of Japan.

\end{document}